\RequirePackage{fix-cm}
\documentclass[twocolumn,epjc3]{svjour3}  %
\smartqed
\RequirePackage{graphicx}
\RequirePackage{amsmath}
\RequirePackage{amsfonts}
\RequirePackage{amssymb}
\RequirePackage{latexsym}
\RequirePackage{widetext}

\newcommand{\be}{\begin{equation}}
\newcommand{\ee}{\end{equation}}
\newcommand{\bea}{\begin{eqnarray}}
\newcommand{\eea}{\end{eqnarray}}

\journalname{Eur. Phys. J. C}
\begin{document}

\title{Probing time orientability of spacetime}

\author{N. A. Lemos\thanksref{e1,addr1}
        \and
				\ D. M\"uller\thanksref{e2,addr2}
				\and
       \  M. J. Rebou\c{c}as\thanksref{e3,addr3}}

\thankstext{e1}{e-mail: nivaldolemos@id.uff.br}
\thankstext{e2}{e-mail: dmuller@unb.br}
\thankstext{e3}{e-mail: reboucas@cbpf.br}

\institute{Instituto de F\'{\i}sica, Universidade Federal Fluminense,
Av. Litor\^anea, S/N, $\,$24210-340 Niter\'oi -- RJ, Brazil\label{addr1}
           \and
Instituto de F\'{\i}sica, Universidade de Bras\'{\i}lia, $\,$
70919-970 Bras\'ilia - DF,  Brazil\label{addr2}					
           \and
Centro Brasileiro de Pesquisas F\'{\i}sicas,
Rua Dr.\ Xavier Sigaud 150,
$\,$22290-180 Rio de Janeiro -- RJ, Brazil\label{addr3}
}

\date{Received: date / Accepted: date}

\maketitle

\sloppy




\begin{abstract}
In general relativity, cosmology and quantum field theory,  spacetime is assumed to be an
orientable  manifold endowed with a Lorentz metric that makes it  spatially and temporally
orientable. The question as to whether the laws of physics require these orientability assumptions
is ultimately of observational or experimental nature, or the answer might come from a fundamental theory of
physics. The possibility that spacetime is time non-orientable lacks investigation, and so should not be dismissed straightaway. 
In this paper, we argue that it is possible to locally access a putative time non-orientability of
Minkowski empty spacetime by physical effects involving quantum vacuum electromagnetic fluctuations.
We set ourselves to study the influence of time non-orientability on the  stochastic motions of a charged particle subject to these electromagnetic
fluctuations in Minkowski spacetime equipped with a time non-orientable topology and with its
time orientable counterpart.
To this end, we introduce and derive analytic expressions for a statistical time orientability
indicator. Then we show that it is possible to pinpoint the time non-orientable topology through an inversion pattern displayed by the  corresponding orientability indicator, which is absent when the underlying manifold is time orientable. 
 \end{abstract}

\PACS{03.70.+k, 05.40.Jc, 42.50.Lc, 04.20.Gz, 98.80.Jk, 98.80.Cq}

\maketitle
\newpage
\section{Introduction} \label{Intro}

In  relativistic cosmology 
and   quantum field theory spacetime is described as a four-dimensional
differentiable manifold,  a topological space with an additional differentiable
structure that permits to locally define  connections, metric and curvature with which
the gravitation theories and the dynamics of other fields  are locally  
formulated. 
So, two important elements in the mathematical description of the Universe
and in the local dynamics of the (micro) physical laws are the geometry and
the topology of the underlying spacetime manifold. 

Geometry is a local attribute that brings about curvature. Topology is
a global property of manifolds that requires consideration of
the entire manifold.
However, if local physics is brought into the scene,
it can play a key role in the local access to the topological
properties of spacetime. This is the main
concern of this work, which focuses on the time orientability of spacetime.

For a manifold endowed with a
Lorentz metric, two possible types of orientability (non-orientability) come
about: spatial or temporal orientability (non-orientability), depending on the
way the manifold is  equipped with a Lorentz metric.
Whether a spacetime is time or
space orientable may be looked upon as a joint topological-geometrical property,
in the sense that it depends on the topology of the underlying manifold but also on
the specific Lorentz metric we equip the manifold
with~\cite{Hawking-Ellis-1973,Geroch-Horowitz-1979,Minguzzi-2019}. 

It is generally assumed that the spacetime manifolds one deals with in physics
are orientable  and  endowed with a
Lorentz metric, making them  separately time and space
orientable.
\footnote{Theoretical arguments in support of these orientability assumptions
combine space and time universality of  local physical laws with the thermodynamically
defined  local arrow of time, charge conjugation and parity (CP) violation and CPT
invariance~\cite{Zeldovich1967,Hawking-Ellis-1973,Geroch-Horowitz-1979}.
The impossibility of having globally defined spinor fields on non-orientable
spacetime manifolds is also often used in favor of these orientability assumptions%
~\cite{Hawking-Ellis-1973,Geroch-Horowitz-1979,Penrose-1968,Penrose-Hindler1986,%
Geroch-1968,Geroch-1970}.
It should be noted, however, that time universality can be looked upon as
a topological assumption of global time homogeneity. 
This topological assumption  rules out time non-orientability from
the outset.}
Non-orientability raises intriguing questions, and it is generally seen as an
undesirable feature in physics. Yet, space and time non-orientability of
Lorentzian spacetime manifolds are concrete mathematical possibilities
for the physical spacetime.  
On the other hand,  strict orientability assumptions could risk ruling out  
something that the spacetime topology might be trying to ``tell'' us,
and it might also discourage further investigations on the interplay of physics
and topology. The answer to questions
regarding the orientability of spacetime might come from local experiments,
cosmological observations or from a fundamental theory
of physics.

In this article we address the question as to whether one can empirically and
locally  access  the putative  topological property of \textsl{time}
non-orientability (orientability) of spacetime manifolds $\mathcal{M}_4$.
\footnote{The nature of time and the existence of various arrows of time
are contested issues, which we evade in this paper, in which we assume that
time is real and passes in all scales. The literature on this debate is
fascinating and vast, but for the sake of brevity we refer the readers to
Refs.~\cite{Ellis-Drosel-2020,Castagnino-Lara-Lombardi-2003} and references
therein, including the books~\cite{Reichenbach-1971,Davies-1974,%
Barbour-2000,Rovelli-2018}.  }
Since the net role played by time orientability  is more
clearly ascertained  in static flat spacetime, whose  dynamical degrees
of freedom are frozen, in this work we  focus on Minkowski spacetime.
The physical system that turns out to be suitable to play the revealing role of
global time non-orientability is a non-relativistic  charged particle
locally subjected to quantum vacuum fluctuations of the electromagnetic field.
\footnote{ In an influential work, this system was used by Yu and Ford~\cite{yf04}
in Minkowski spacetime with a conducting plane (nontrivial spatial topology). They
endeavored to shed  light on the question as to whether a test charged particle
would perform stochastic motions  induced by quantum fluctuations of the electromagnetic
field~\cite{gs99,jr92}. Related investigations have since been made in a number of papers,
including
Refs.~\cite{Bessa:2019aar,Yu:2004gs,Ford:2005rs,Yu:2006tn,Seriu:2008,%
Parkinson:2011yx,DeLorenci:2016jhd,Lemos:2020ogj,Lemos-Muller-Reboucas-2022}. }

To show that one can access time orientability,  we investigate signatures
of time non-orientability  through  the stochastic motions of a charged test
particle under electromagnetic quantum fluctuations in Minkowski
spacetime $\mathcal{M}_4$ with a time non-orientable  topology and in  its
time orientable counterpart. The time orientability statistical
indicator that we introduce combines geometrical-topological properties with the
dynamics of the above specific physical system engineered to test orientability.

The structure of the paper is as follows.
In Section~\ref{TopSet} we set up the notation and present some key concepts
and results regarding topology of manifolds, which will be needed in the
remainder of the paper.
In Section~\ref{Sec-charge} we present the physical system along with the background
geometry and topology. In Section \ref{Mobius} we introduce the time orientability
statistical indicator   and derive its expressions for a  charged particle under
quantum vacuum electromagnetic fluctuations in Minkowski spacetime equipped with
temporally non-orientable and orientable topologies.
We show that a comparison of the time evolution of our statistical indicators
for these cases allows one to discriminate the orientable from the non-orientable
topology.
Time non-orientability can be locally  unveiled by the inversion pattern
of the curves of the time orientability statistical indicator for a point charge
under quantum vacuum electromagnetic fluctuations. In
Section~\ref{Conclusions} we present our main conclusions and final
remarks.

\section{Context and mathematical preliminaries}  \label{TopSet}  

For both the spatially flat Friedmann-Robertson-Walker (FRW) and the
 Minkowski spacetimes, the underlying  manifold $\mathcal{M}_4$
is  globally decomposable as $\mathcal{M}_4 = \mathbb{R}\times M^3$,
where $M^3$ is the simply-connected  $3$-manifold $\mathbb{R}^{3}$.
However, the spatial section $M^3$ of both spacetimes can be any quotient
manifold of the form $M^3=\mathbb{E}^{3}/\Gamma$, where $\mathbb{E}^{3}$ is
the covering space,\footnote{$\mathbb{E}^n$ is $\mathbb{R}^n$ endowed with the
Euclidean metric.} and $\Gamma$ is a discrete  group of freely acting
isometries of $\mathbb{E}^{3}$, also referred to as the  
holonomy~\cite{Wolf67,Thurston}.$\,$%
\footnote{Refs.~\cite{Adams-Shapiro01,Cipra02,Riazuelo-et-el03,Fujii-Yoshii-2011}
give a detailed account of the classification of flat $3$-dimensional
Euclidean topologies.}
The familiar form  $\mathcal{M}_4 = \mathbb{R}\times M^3$ is very often assumed
in cosmology and quantum field theory, and in particular it is adopted 
in cosmic topology, which investigates the spatial topology of the Universe.
\footnote{For investigations on cosmic topology and recent observational constraints, see  Refs.~\cite{Ellis:1970ey,LACHIEZEREY1995135,Starkman:1998qx,LEVIN2002251,Reboucas:2004dv,%
Reboucas-2005,Luminet:2016bqv,2014,2016,Cornish:2003db,ShapiroKey:2006hm,Bielewicz:2010bh,%
Vaudrevange:2012da,Aurich:2013fwa,Gomero:2016lzd}.}

In the present work we focus on a single topological
property of the Minkowski spacetime manifold $\mathcal{M}_4$,
namely \textsl{time orientability}. 
Perhaps the simplest way to equip  Minkowski spacetime  with 
time non-orientability is by taking 
\begin{equation}  \label{Min_M2-R2}
\mathcal{M}_4 =  \mathcal{M}^2 \times  \mathbb{E}^2 \,,
\end{equation}
where  $\mathcal{M}^2$ is a two-dimensional time non-orientable quotient
manifold, $\mathcal{M}^2 = \mathbb{E}^{2,1}/ \Gamma$, 
where $\mathbb{E}^{2,1}$ denotes the simply-connected  plane equipped with a
Lorentz metric, that is, Minkowski two-dimensional spacetime.
It follows that the orientability of $\mathcal{M}_4$ reduces to the
orientability of  $\mathcal{M}^2\,$.%

The simplest example of a two-dimensional Euclidean manifold with nontrivial
topology is the cylinder $\mathbb{C}^2$, whose construction as a quotient
manifold, $\mathbb{C}^2 =\mathbb{E}^2/\Gamma$, is such that a point $P=(x,y)$ of
the cylinder is obtained from the covering manifold $\mathbb{E}^2$
by identifying the points that are equivalent under the action of the elements $\gamma_i$
of the covering isometry group $\Gamma$.
Specifically,  a point $P=(x,y)$ of the cylinder is the equivalence class of all points $P'$
in the covering space $\mathbb{E}^2$ such that 
$ P' = \{( x + n_x a, y) \mid  n_x \in  \mathbb{Z}, a = {\rm const}\}$,
or $P  \equiv  P' = \gamma_i P $ with $\gamma_i \in \Gamma $
being a translation by $a$ in the direction of the $x$-axis.
A   quotient manifold can be visualized
by its so-called fundamental domain (cell). The cylinder's  fundamental cell
is a strip of $\mathbb{E}^2$ bounded by parallel lines, say $x=0$ and
$x=a$, that are identified through translations.
The cylinder $\mathbb{C}^2$ is  a surface
lying in three-dimensional space, but this embedded view  does not
necessarily work for other Euclidean two-manifolds.

The twisted cylinder $\mathbb{C^*}^2$ is an example of manifold
that cannot lie in ordinary three-dimensional space without intersecting
itself~\cite{Stillwell-1992}.
A point $P=(x,y)$ of  $\mathbb{C^*}^2$ represents
a set of points in  the covering space $\mathbb{E}^2$ of the form
$ P' =\{( x + n_x a , (-1)^{n_x} y) \mid n_x \in  \mathbb{Z},
\,a={\rm const} \}$
or $P \equiv P' = \gamma_i P $ with $\gamma_i \in \Gamma $
being  translation by $a$ in the $x$-direction followed by an inversion
(flip) in the $y$-direction, a single glide reflection. The fundamental
cell of the twisted cylinder is a strip of $\mathbb{E}^2$ bounded by parallel lines
which are identified through a glide reflection: translation followed by a flip
(inversion).
It should be noticed that the twisted cylinder is often represented by the M\"obius strip, 
which is obtained by identifying two opposite sides of a rectangle after a flip. 
The twisted cylinder is an infinitely wide M\"obius strip. The M\"obius strip is visually 
useful as it can lie in $3-$space, but it is not a manifold because it has a boundary. 
The twisted cylinder, on the other hand, is a genuine quotient manifold but cannot 
lie in $3-$space.

At this point it seems fitting to remark that a quotient manifold is globally
homogeneous only if its fundamental cells are identified by translations alone.
Therefore, the cylinder $\mathbb{C}^2$ is globally homogeneous but the twisted
cylinder $\mathbb{C^*}^2$ is not.

Orientability is another very important global (topological) property of
a manifold that measures whether one can  consistently choose a clockwise
orientation for loops in the manifold.
A path in   $\mathcal{M}^2$ that brings a traveler back to the starting point
mirror-reversed is called an orientation-reversing path. Manifolds that do not
have an orientation-reversing path  are called
\textsl{orientable}, whereas
manifolds that contain an orientation-reversing path
are \textsl{non-orientable}~\cite{Weeks2020}.
For two-dimensional quotient manifolds $\mathbb{E}^2/\Gamma$, when the
covering group $\Gamma$ contains at least one isometry $\gamma$ that
is a reflection (flip) the corresponding quotient manifold is
non-orientable. Therefore, the cylinder is orientable but the twisted
cylinder is non-orientable.

For the product manifold given by equation \eqref{Min_M2-R2} to be
time non-orientable the factor $\mathcal{M}^2$ has to be a Lorentzian
non-orientable manifold. It turns out that both the cylinder and the
twisted cylinder can be equipped with a Lorentz metric
--- see Ref.~\cite{ONeill-1983}, p. 149, Proposition 37.
The twisted cylinder is made time non-orientable by endowing it with
the Lorentz metric with time as the flipped direction.


\section{Charged particle under electromagnetic fluctuations}
\label{Sec-charge}

From now on, we consider a point charge under quantum vacuum electromagnetic
fluctuations as the physical system used to locally probe a potential time
non-orientablity of spacetime.

\subsection{The physical system}

Let a  nonrelativistic test particle with
charge $q$ and mass $m$ be locally subject to
vacuum fluctuations of the electric field ${\bf E}({\bf x}, t)$
in a topologically nontrivial spacetime manifold
equipped with the Minkowski metric $\eta_{\mu\nu}=\mbox{diag} (+1, -1, -1, -1)$.

Locally, the motion of the  charged test particle is determined by the Lorentz force.
In the nonrelativistic limit  the equation of motion for the point charge is
\begin{equation}\label{eqmotion1}
\frac{d{\bf v}}{dt} = \frac{q}{m} \,{\bf E}({\bf x}, t)\,,
\end{equation}
where  $\mathbf{v}$ is the particle's velocity and
$\mathbf{x}$ its position at  time $t$.
We assume that on the time scales of interest the particle practically does not
move, i.e. it has a negligible displacement, so we  can ignore the time
dependence of  $\mathbf{x}$.
Thus, the  particle's position $\mathbf{x}$ is taken as constant in what
follows~\cite{yf04}.%
\footnote{The corrections arising from the inexactness of this assumption
are negligible in the low velocity regime.}
Assuming that the particle is initially  at rest,  integration of
Eq.~\eqref{eqmotion1} gives
\begin{equation}\label{eqmotion2}
{\bf v}({\bf x}, t) = \frac{q}{m}\int_{t_0}^{t}{\bf E}({\bf x}, t^{\prime})
\,dt^{\prime} \,,
\end{equation}
and the mean squared velocity, velocity dispersion or simply dispersion in each of
the three independent directions $i = x, y, z$ is given by%
\footnote{By definition,
$\,\bigl \langle \,\Delta v^2_i(\mathbf{x}, t) \,\bigr \rangle =
\bigl \langle \, v^2_i(\mathbf{x}, t)  \,\bigr \rangle
-\bigl \langle \,v_i(\mathbf{x}, t)   \bigr \rangle^2$.}
\begin{equation}\label{eqdispersion1}
\Bigl \langle\Delta v^2_i \Bigr\rangle = \frac{q^2}{m^2} \int_{t_0}^t\int_{t_0}^t
\Bigl\langle E_i({\bf x}, t') E_i({\bf x}, t'')\Bigr \rangle\, dt' dt''\,.
\end{equation}
Following Yu and Ford~\cite{yf04}, we assume that the electric field is
a sum of classical  $\mathbf{E}_c$ and quantum  $\mathbf{E}_q$ parts.
Because ${\bf E}_c$ is not subject to quantum fluctuations and
$\langle {\bf E}_q\rangle =0$,
the two-point function $\langle E_i({\bf x}, t)E_i({\bf x}', t')\rangle$ in
equation~\eqref{eqdispersion1} involves only
the quantum part of the electric field~\cite{yf04}.

It can be shown~\cite{bd82} that locally
\begin{eqnarray} \label{eqdif-0}
\Bigl \langle E_i({\bf x}, t)E_i({\bf x}', t') \Bigr \rangle & = &
\frac{\partial }{\partial x_i} \frac{\partial}
{\partial {x'}_i}D({\bf x}, t; {\bf x}', t') \nonumber \\
& &   -  \frac{\partial }{\partial t}
 \frac{\partial}
{\partial t'}D ({\bf x }, t; {\bf x'}, t')
\end{eqnarray}
where, in Minkowski spacetime,
the Hadamard function $D({\bf x}, t;{\bf x}', t')$ is given by
\begin{equation}\label{eqren}
D_0({\bf x}, t; {\bf x}', t') = \frac{1}{4\pi^2(\Delta t^2 - |\Delta \mathbf{x}|^2)} \,.
\end{equation}
The subscript $0$ indicates standard Minkowski spacetime $\mathbb{R} \times \mathbb{E}^3$,
$\Delta t = t - t'$ and  $|\Delta \mathbf{x}| \equiv r $  
is the spatial separation for topologically trivial Minkowski spacetime:
\begin{equation}\label{separation-trivial}
r^2 = (x-x')^2 + (y-y')^2 + (z - z')^2  \,.
\end{equation}

\subsection{The spacetime manifold}
\begin{table}[b]
\centering
\caption{Time  and space separation for the spacetime interval $\Delta s^2 = \Delta t^2 - r^2$
in the Hadamard function for the spacetime manifolds obtained by taking the Cartesian product
with $\mathbb{E}^2$ of  the two-dimensional manifolds $\mathbb{C}^{*2}$ or $\mathbb{C}^{2}$
with coordinates $t,x$;  the coordinates associated with  $\mathbb{E}^2$ are $y,z$.
With time always in the first factor, each flat four-manifold thus obtained  is endowed with
a Lorentz metric with signature $-2$. In the case of $\mathbb{C}^{*2}$  the identification
with inversion is made on the time direction, so that the corresponding spacetime is not
time orientable. The  topological compact spatial scale is denoted by $a$.
The number $n_x$ is an integer  that runs from $-\infty$ to $\infty$. } %
\label{Tb-Time-Spatial-separation}
\begin{tabular*}{\columnwidth}{l|c|c}
\hline
\hspace*{-3mm} Manifold                               & \hspace*{-2.5mm} Time sep. $\Delta t$ &  Spatial sep. $r^2$ \\
\hline
\hspace*{-2mm}\mbox{\footnotesize $\mathbb{C}^{*2} \times \mathbb{E}^2$}  & \hspace*{-2.5mm} \mbox{\scriptsize $t - (-1)^{n_x} t^{\prime}$} & \hspace*{-2.4mm}\mbox{\scriptsize $(x-x'-n_x a)^2 + (y - y')^2 + (z - z')^2$} \\
  \hline
\hspace*{-2.5mm} \mbox{\footnotesize $\mathbb{C}^2\times \mathbb{E}^2$}  & \hspace*{-2.5mm} \mbox{\scriptsize$t - t^{\prime}$}     & \hspace*{-2.5mm} \mbox{\scriptsize $(x - x'- n_x a)^2 + (y - y')^2 + (z - z')^2 $} \\
\hline
\end{tabular*}
\end{table}
In this work, the  time non-orientable spacetime manifold that we shall consider is of
the form $\mathcal{M}_4 =  \mathcal{M}^2 \times  \mathbb{E}^2$ in which the first factor
$\mathcal{M}^2$ is the non-orientable twisted cylinder $\mathbb{C}^{*2}$ equipped with a
Lorentz metric with time as the flipped coordinate. This makes $\mathcal{M}^2$ \textsl{time}
non-orientable and, as a consequence, the spacetime manifold $\mathcal{M}_4$ is also time
non-orientable.
In Table \ref{Tb-Time-Spatial-separation} we show the time separation $\Delta t$ and the
spatial separation $r$ that enter the spacetime interval $\Delta s^2 = \Delta t^2 - r^2$ in
$\mathcal{M}_4$ for  $\mathcal{M}^2 = \mathbb{C}^{*2}$ and $\mathcal{M}^2 = \mathbb{C}^{2}$,
the cases  treated in this paper.  For $\mathcal{M}^2 = \mathbb{C}^{*2}$ the non-orientability
is associated with the time coordinate.


Let us try to make our terminology and notation as clear as possible. Being aware of the
abuse of language but striving for simplicity,  we henceforward give the name  {\it twisted cylinder},
denoted by $\mathbb{C}^{*2} \times \mathbb{E}^2$, to the time non-orientable spacetime
with coordinates $(t,x,y,z)$, whose points are identified as $(t,x,y,z) \equiv ((-1)^{n_x}t, x+n_x a,y,z)$,
and equipped with the Lorentz metric $\Delta s^2 = \Delta t^2- r^2$, where $\Delta t$ and
$r$ are given in the first line of Table \ref{Tb-Time-Spatial-separation}.  Similarly,
we give the name {\it  cylinder}, denoted by $\mathbb{C}^{2} \times \mathbb{E}^2$, to the
time orientable spacetime  with coordinates $(t,x,y,z)$, whose points are identified as
$(t,x,y,z) \equiv (t, x+n_x a,y,z)$, and equipped with the Lorentz metric $\Delta s^2 = \Delta t^2
- r^2$, where $\Delta t$ and $r$ are given in the second line of Table \ref{Tb-Time-Spatial-separation}.

\section{TIME-ORIENTABILITY INDICATOR FOR THE TWISTED CYLINDER}
\label{Mobius}

We take up now the study of the stochastic motions of a charged particle under
quantum vacuum electromagnetic fluctuations in the  time non-orientable
twisted cylinder $\mathbb{C}^{*2} \times \mathbb{E}^2$.

In a topologically nontrivial spacetime, the time interval $\Delta t$ and the
the spatial separation $r$ take  new forms that capture the
periodic boundary conditions imposed on the covering space.
To obtain the correlation function for the electric field
that is required to compute the velocity dispersion~\eqref{eqdispersion1} for
the twisted cylinder $\mathbb{C}^{*2} \times \mathbb{E}^2$,  we replace in
Eq.~(\ref{eqdif-0}) the Hadamard
function $D({\bf x}, t; {\bf x}', t')$ by its renormalized version
given by~\cite{Bessa:2019aar}
\begin{eqnarray}\label{Hadamard-ren}
D_{ren}({\bf x}, t; {\bf x}', t') & = &
D({\bf x}, t; {\bf x}', t') - D_0({\bf x}, t; {\bf x}', t') \nonumber \\
& = & \sum\limits_{{n_x=-\infty}}^{{\infty\;\;\prime}}\frac{1}{4\pi^2(\Delta t^2 - r^2)}\,,
\end{eqnarray}
where  here and in what follows $\sum_{}^{\;'}$ indicates that
the Minkowski contribution term $n_x = 0$ is excluded from the summation,
and, according to Table \ref{Tb-Time-Spatial-separation},
the time and space separations for the time non-orientable spacetime
$\mathbb{C}^{*2} \times \mathbb{E}^2$ are
\begin{equation}
\label{separation-slab-time-non-orient}
\Delta t  =  t-(-1)^{n_x}t^{\prime}, \, r^2 = ( x - x' - n_x a)^{2} +(y -y')^2
+ (z - z')^{2}.
\end{equation}
The infinite sum \eqref{Hadamard-ren} takes account of the periodicity induced by the cell
identification that defines $\mathbb{C}^{*2} \times \mathbb{E}^2$ in terms of the covering space.
This approach is widely used~\cite{Dowker-Critchley-1976,Dowker-Banach-1978,dhi79,Gomero-etal,st06,DHO-2001,DHO-2002,MD-2007,Matas}
and does not mean that local physics is influenced by arbitrarily distant regions that are causally
disconnected from the region of interest because the covering space (where the calculations are performed)
is not the physical space.
The term  with $n_x=0$ in  the sum~\eqref{Hadamard-ren} is the Hadamard function
$D_{0}({\bf x}, t; {\bf x}', t')$ for  Minkowski spacetime.
This term  has been subtracted out from the sum
because it gives rise to an infinite contribution
to the velocity dispersion.  

Thus, from equation \eqref{eqdif-0} the renormalized correlation functions
\begin{eqnarray}\label{correlation-general}
\bigl \langle E_i({\bf x}, t)E_i({\bf x}', t')\bigr \rangle_{ren} & = &
\frac{\partial }{\partial x_i} \frac{\partial}
{\partial {x'}_i}D_{ren}({\bf x}, t; {\bf x}', t') \nonumber \\
& &-\frac{\partial }{\partial t} \frac{\partial}
{\partial t'}D_{ren} ({\bf x }, t; {\bf x'}, t')
\end{eqnarray}
are then given by
\begin{eqnarray}
\label{correlation-i-slab-time-non-orient}
 \bigl \langle E^i({\bf x}, t)E^i({\bf x}', t')\bigr \rangle_{ren}^{\mathbb{C}^{*2}} & = & \sum\limits_{{n_x=-\infty}}^{{\infty\;\;\prime}}
\bigg\{ \frac{[3(-1)^{n_x} -1]\Delta t^2 }{2\pi^2 [\Delta t^2 - r^2]^3} \nonumber \\
& & + \frac{ [1+ (-1)^{n_x}]r^2 -4r_i^2}{2\pi^2 [\Delta t^2 - r^2]^3}\bigg\},
\end{eqnarray}
where $\Delta t$ and  $r^2$ are given by Eq. \eqref{separation-slab-time-non-orient}, while
\begin{equation}
\label{ri}
r_1 \equiv r_x = x- x^{\prime}-n_xa, \,\, r_2 \equiv r_y = y- y^{\prime},
\,\, r_3 \equiv r_z = z- z^{\prime}.
\end{equation}

The orientability indicator $\mbox{\large $I$}_{v^2_i}^{\mathbb{C}^{*2}}$ that
we will consider
is defined by replacing
the electric field correlation functions in Eq.~(\ref{eqdispersion1}) by
their renormalized counterparts:\footnote{To avoid a cluttered notation,
for the superscript  on the statistical indicator and on the electric field
correlation functions we write just $\mathbb{C}^{*2}$ instead of
$\mathbb{C}^{*2} \times \mathbb{E}^2$.}
\begin{equation}   \label{indicator-TNS}
\mbox{\large $I$}_{v^2_i}^{\mathbb{C}^{*2}} ({\bf x},t,t_0) = \frac{q^2}{m^2} \int_{t_0}^t\int_{t_0}^t
\Bigl\langle E^i({\bf x}, t') E^i({\bf x}, t'')\Bigr \rangle_{ren}^{\mathbb{C}^{*2}}\, dt' dt''\,.
\end{equation}
From~(\ref{Hadamard-ren}) it is clear  that the orientability indicator
$\mbox{\large $I$}_{v^2_i}^{\mathbb{C}^{*2}}$ is the difference between the velocity
dispersion in $\mathbb{C}^{*2} \times \mathbb{E}^2$ and the one in Minkowski spacetime with trivial
topology.

Before moving on to explicit computations,  a few words of clarification on the meaning
of the indicator~\eqref{indicator-TNS} are fitting.
From equations~\eqref{eqdispersion1} and~\eqref{Hadamard-ren}
a general definition of the orientability indicator can be written in the form
\begin{equation}  \label{new-ind}     
 \mbox{\large $I$}_{v^2_i}^{MC}
= \Bigl\langle\Delta v_i^2 \Bigr \rangle^{MC}
- \;\,\Bigl \langle\Delta v_i^2 \Bigr \rangle^{SC} ,
\end{equation}
where  $\bigl\langle\Delta v_i^2 \bigr \rangle$  is the
mean square velocity dispersion, and the superscripts $MC$ and $SC$
stand for multiply- and simply-connected manifolds, respectively.
The right-hand side of~(\ref{new-ind}) is defined by first taking the difference
of the two terms with ${\bf x}^{\prime} \neq  {\bf x}$ and then setting
${\bf x}^{\prime} =  {\bf x}$.

On the face of it, the indicator \eqref{new-ind} does not appear to be measurable
because it involves the difference of quantities associated with
two different spacetimes, but the  spacetime we live in is unique.
However,  $\mbox{\large $I$}_{v^2_i}^{MC}$ is accessible by measurements 
performed in our  spacetime, which is to be tested for time non-orientablity, by the
following procedure. First one would measure the  velocity correlation function
$\bigl\langle \Delta v_i({\bf x},t) \Delta v_i({\bf x}^{\prime},t)\bigr\rangle^{MC}$ for
${\bf x} \neq {\bf x}^{\prime}$, then  one would subtract out the correlation
function $\bigl\langle \Delta v_i({\bf x},t) \Delta v_i({\bf x}^{\prime},t)\bigr\rangle^{SC}$
that has been {\it  theoretically computed}  for ${\bf x} \neq {\bf x}^{\prime}$
for the corresponding  topologically trivial Minkowski spacetime in the Appendix of
Ref.~\cite{Lemos:2020ogj}.
Finally, the corresponding curve for the difference \eqref{new-ind} as a function of
time would be plotted in the coincidence limit ${\bf x} = {\bf x}^{\prime}$.
\footnote{This approach is analogous to one used in the search for spatial topology
of the universe from  discrete cosmic sources, called cosmic crystallography~\cite{LeLaLu},
in which a  topological signature of  $3-$space is given by a constant times the
difference $\Phi_{exp}^{MC}(s_i) - \Phi^{SC}_{exp}(s_i)$ of the
expected pair separation histogram (EPSH), and the EPSH for the
underlying simply connected covering manifold~\cite{GRT01,GTRB98},
which  can be theoretically computed in analytical
form~\cite{GRT01,Reboucas-2000}.}

\subsection{Indicators for the twisted cylinder}

The orientability indicator $\mbox{\large $I$}_{v^2_i}^{\mathbb{C}^{*2}}$
can  be computed with
the help of the integrals~\cite{Bessa:2019aar}
\begin{eqnarray}\label{integral1}
I_- & =& \int_{t_0}^t  \int_{t_0}^t dt^{\prime}dt^{\prime\prime}
\frac{1}{[(t^\prime-t^{\prime\prime})^2  -r^2]^3} \nonumber\\
 &= & \frac{3(t-t_{0})}{16r^{5}}\ln\left[\left(\frac{t_{0}-t+r}{t_{0}-t-r}\right)^{2}\right]\nonumber \\
 & & +\frac{1}{4r^{2}\left[(t-t_{0})^{2}-r^{2}\right]}
+\frac{1}{4r^{4}}
\end{eqnarray}
and
\begin{eqnarray}\label{integral2}
J_- & = &\int_{t_0}^t  \int_{t_0}^t dt^{\prime} dt^{\prime\prime}
\frac{(t^\prime-t^{\prime\prime})^2}{[(t^\prime-t^{\prime\prime})^2 -r^2]^3}  \nonumber \\
& = &\frac{(t-t_{0})}{16\,r^{3}}\ln\left[\left(\frac{t_{0}-t-r}{t_{0}-t+r}\right)^{2}\right] \nonumber \\
& & +\frac{1}{4\left[(t-t_{0})^{2}-r^{2}\right]}
+\frac{1}{4r^{2}}
\end{eqnarray}
as well as
\begin{eqnarray}\label{integral-prime1}
I_+ & = &\int_{t_0}^t  \int_{t_0}^t dt^{\prime}dt^{\prime\prime}\frac{1}{[(t^\prime+t^{\prime\prime})^2  -r^2]^3} \nonumber \\
& = & \frac{3t_{0}}{16r^{5}}\ln\left[\left(\frac{2t_{0}-r}{2t_{0}+r}\right)^2\left(\frac{t_{0}+t+r}{t_{0}
 +t-r}\right)^2\right]    \nonumber \\
& & +\frac{1}{8r^{2}\left[4t_{0}^{2}-r^{2}\right]}\nonumber\\
& & +\frac{3t}{16r^{5}}\ln\left[\left(\frac{t_{0}+t+r}{t_{0}+t-r}\right)^2\left(\frac{2t-r}{2t+r}\right)^2\right] \nonumber \\
& &  + \frac{1}{8r^{2}\left[4t^{2}-r^{2}\right]}-\frac{1}{4r^{2}\left[(t+t_{0})^{2}-r^{2}\right]}
\end{eqnarray}
and
\begin{eqnarray}\label{integral-prime2}
 J_+ & = &\int_{t_0}^t  \int_{t_0}^t dt^{\prime} dt^{\prime\prime}
\frac{(t^\prime+t^{\prime\prime})^2}{[(t^\prime+t^{\prime\prime})^2 -r^2]^3} \nonumber \\
& = & \frac{t_{0}}{16r^{3}}\ln\left[\left(\frac{2t_{0}+r}{2t_{0}-r}\right)^2\left(\frac{t_{0}+t-r}{t_{0}+t+r}\right)^2\right]
\nonumber \\
& & +\frac{1}{8\left[4t_{0}^{2}-r^{2}\right]}\nonumber\\
& & +\frac{t}{16r^{3}}\ln\left[\left(\frac{t_{0}+t-r}{t_{0}+t+r}\right)^2\left(\frac{2t+r}{2t-r}\right)^2\right] \nonumber \\
& &  +\frac{1}{8\left[4t^{2}-r^{2}\right]}-\frac{1}{4\left[(t+t_{0})^{2}-r^{2}\right]}
\end{eqnarray}

By using these integrals and equations~(\ref{correlation-i-slab-time-non-orient})
and \eqref{ri}  in Eq.~(\ref{indicator-TNS}) we find
\begin{eqnarray}\label{indicator-i-TNS}
\mbox{\large $I$}_{v^2_i}^{\mathbb{C}^{*2}}({\bf x},t,t_0)  & = &
\frac{q^2}{\pi^2m^2}\bigg\{ \sum_{\mbox{\scriptsize even} \,\, n_x}^{\hspace{.4cm} \prime} \bigl[ J_-
+(r^2-2r_i^2)I_-\bigr] \nonumber \\
& & \hspace*{12mm} -2 \sum_{\mbox{\scriptsize odd} \,\, n_x} \bigl[ J_+ + r_i^2I_+\bigr] \bigg\}.
\end{eqnarray}
where, since the coincidence limit ${\bf x} = {\bf x}^{\prime}$ has been taken,
\begin{equation}
\label{r-ri-coincidence-r}
r_1= -n_x a, \,\, r_2 = r_3 =0, \,\, r^2 = r_1^2+r_2^2 +r_3^2 = n_x^2 a^2,
\end{equation}
as follows from Eq.~(\ref{ri}) in the coincidence limit.

\subsection{Indicators for the cylinder}
We are looking for a local way to probe a putative time non-orientability of spacetime.
To this end, let us compare the above results for the twisted cylinder $\mathbb{C}^{*2}
\times \mathbb{E}^2$ manifold with those for its  time-orientable counterpart,
the cylinder
$\mathbb{C}^{2} \times \mathbb{E}^2 =  \mathbb{R} \times \mathbb{S}^1 \times \mathbb{E}^2$.
The indicators for the cylinder
are given in Ref. \cite{Lemos:2020ogj} as\footnote{In equations (26) and (27) of
Ref.~\cite{Lemos:2020ogj} the choice $t_0=0$ was made. Also, what we call here
the cylinder is denoted there by $E_{16}$.}
\begin{eqnarray}
\label{IE16x}
& & I^{\mathbb{C}^2}_{v_x^2}({\bf x},t,t_0)  =  -\frac{q^2 (t-t_0)}{4\pi m^2}\sum_{n_x}^\prime
\frac{1}{r^3}\ln \frac{(r-t+t_0)^2}{(r+t-t_0)^2},\\
& & I^{\mathbb{C}^2}_{v_y^2}({\bf x},t,t_0)=I^{\mathbb{C}^2}_{v_z^2}({\bf x},t)  = \frac{q^2 (t-t_0)}{8\pi m^2} \nonumber \\
& & \hspace*{.2cm} \times \sum_{n_x}^\prime\bigg[\frac{4(t-t_0)}{r^2[(t-t_0)^2-r^2]}  +\frac{1}{r^3}\ln \frac{(r-t+t_0)^2}{(r+t-t_0)^2}\bigg],
\label{IE16y}
\end{eqnarray}
where $r =  n_x a$.

\subsection{Locally probing time non-orientability}

The time-inversion scale is specified by the
topological length scale $a$.
In general, the parameter $a$ leaves open the scale of time-non-orientability,
whose local manifestation is captured by our indicator \eqref{indicator-TNS}.
For example, the parameter $a$ can be very small (microscopic) or very large
(cosmological). Our calculations hold regardless of its value.

\begin{figure*}[htpb]
 \begin{center}
\begin{tabular}{c c}
\resizebox{7.5cm}{!}{\includegraphics{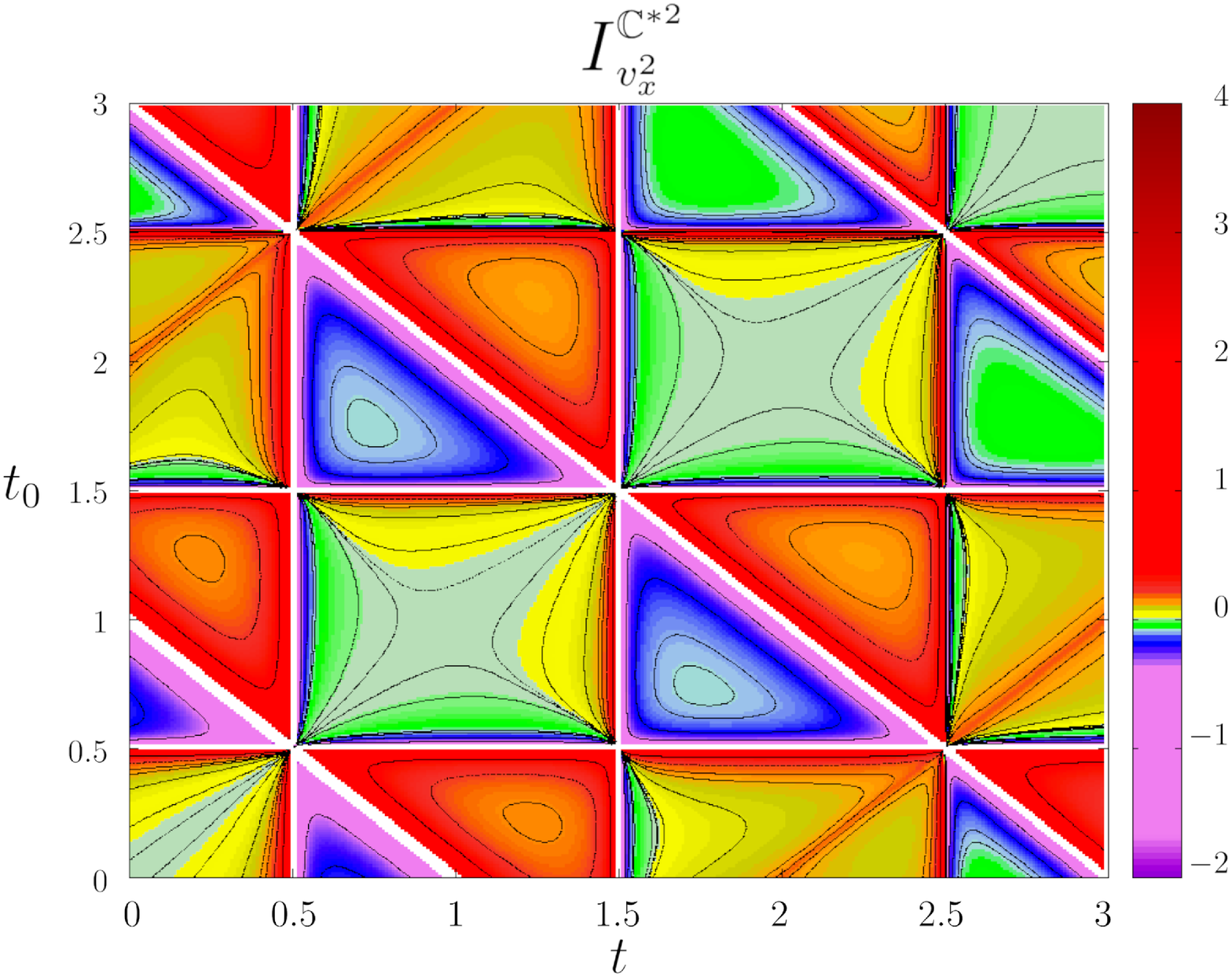}}
    &    \resizebox{7.5cm}{!}{\includegraphics{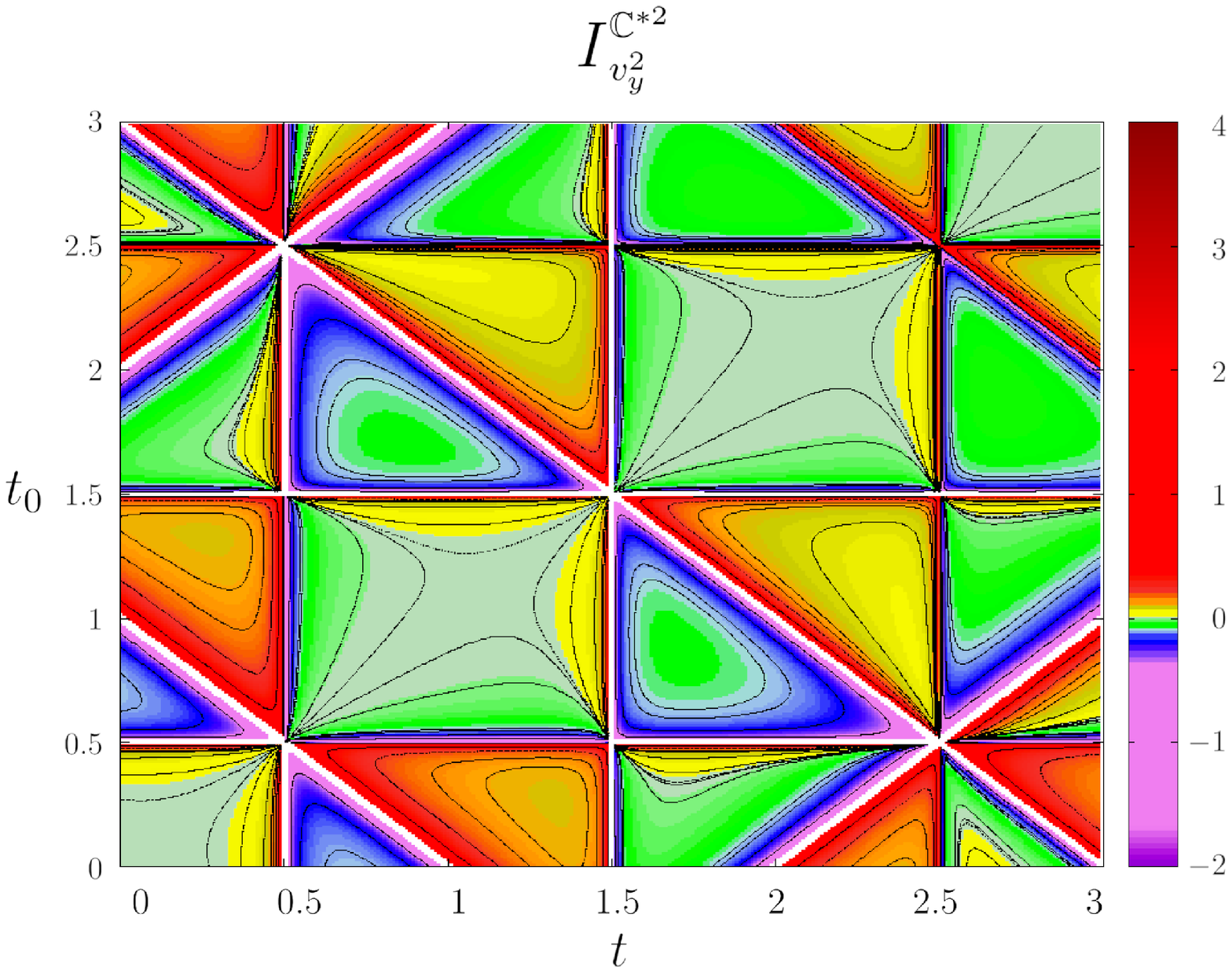}}  \\
       (a) & (b)
  \end{tabular}
    \end{center}
    \caption{Two components of indicator  \eqref{indicator-i-TNS} as functions of $t,t_0$.
    The values of the indicator are represented by colors. Panel (a) shows the  $x$-component
    of the indicator \eqref{indicator-i-TNS}, while panel (b) shows its $y$-component
    (the $z$-component is not displayed because it equals the $y$-component). Contour lines of
    the indicators are also exhibited.  In these plots, $q,\,m$ and $a$ are all set to unity
    for qualitative reasons only. The white thin lines are topological singularities that arise
    for special values of $t, t_0$, such as $t-t_0=r$, $t+t_0 = r$, $t=r/2$, $t_0=r/2$, where
    $r=\vert n_x \vert a$.  The structure of  singularities of topological origin is much
    richer than the one in the case of time-orientable spacetimes~\cite{Bessa:2019aar,%
    Lemos:2020ogj,Lemos-Muller-Reboucas-2022}.
\label{Fig3}}
\end{figure*}

The series \eqref{indicator-i-TNS}, \eqref{IE16x} and  \eqref{IE16y} are rapidly
convergent but we are unable to sum them in closed form. Thus, for the numerical
calculations that allow us to produce some figures, we truncate them
at $\vert n_x \vert = 50$.

A significant difference between the case of the time non-orientable twisted cylinder
and that of time-orientable spacetimes is that the  indicator~\eqref{indicator-i-TNS}
depends both on $t$ and $t_0$, and not only on the difference $t-t_0$. This is because
the twisted cylinder $\mathbb{C}^{*2} \times \mathbb{E}^2$ is not globally temporally
homogeneous.
By means of a picture, we highlight the main consequences of the twisted cylinder's lack of
global time homogeneity on the statistical indicator~\eqref{indicator-i-TNS}.
In Fig.~\ref{Fig3} we  show  two components of indicator \eqref{indicator-i-TNS} as functions
of $t,t_0$, with the indicator values represented by colors (the $z$-component of the indicator
is not shown because it  coincides with its $y$-component). Contour lines of the indicator are
also displayed.  Figs.~\ref{Fig3} (a) and (b) illustrate the  global temporal inhomogeneity
of the twisted cylinder $\mathbb{C}^{*2} \times \mathbb{E}^2$: the patterns encountered as
one moves horizontally along the $t$-axis  are different for each fixed $t_0$.
The nontrivial contour curves are also very different from those for $\mathbb{C}^2 \times \mathbb{E}^2$,
which are the straight lines $t- t_0 = \mbox{const}$. There are periodic repetition patterns but with
changing  absolute value of the indicators. The topological singularity structure, indicated by the
white straight lines, is also much richer than the one for the time orientable case.
For example, the vertical lines $t = 1/2, 3/2, 5/2, \ldots $ are singularities of topological
nature that are not present in the indicators for the time orientable spacetime
$\mathbb{C}^{2} \times \mathbb{E}^2$.

\begin{figure*}[htpb]
 \begin{center}
\begin{tabular}{c c}
\resizebox{7cm}{!}{\includegraphics{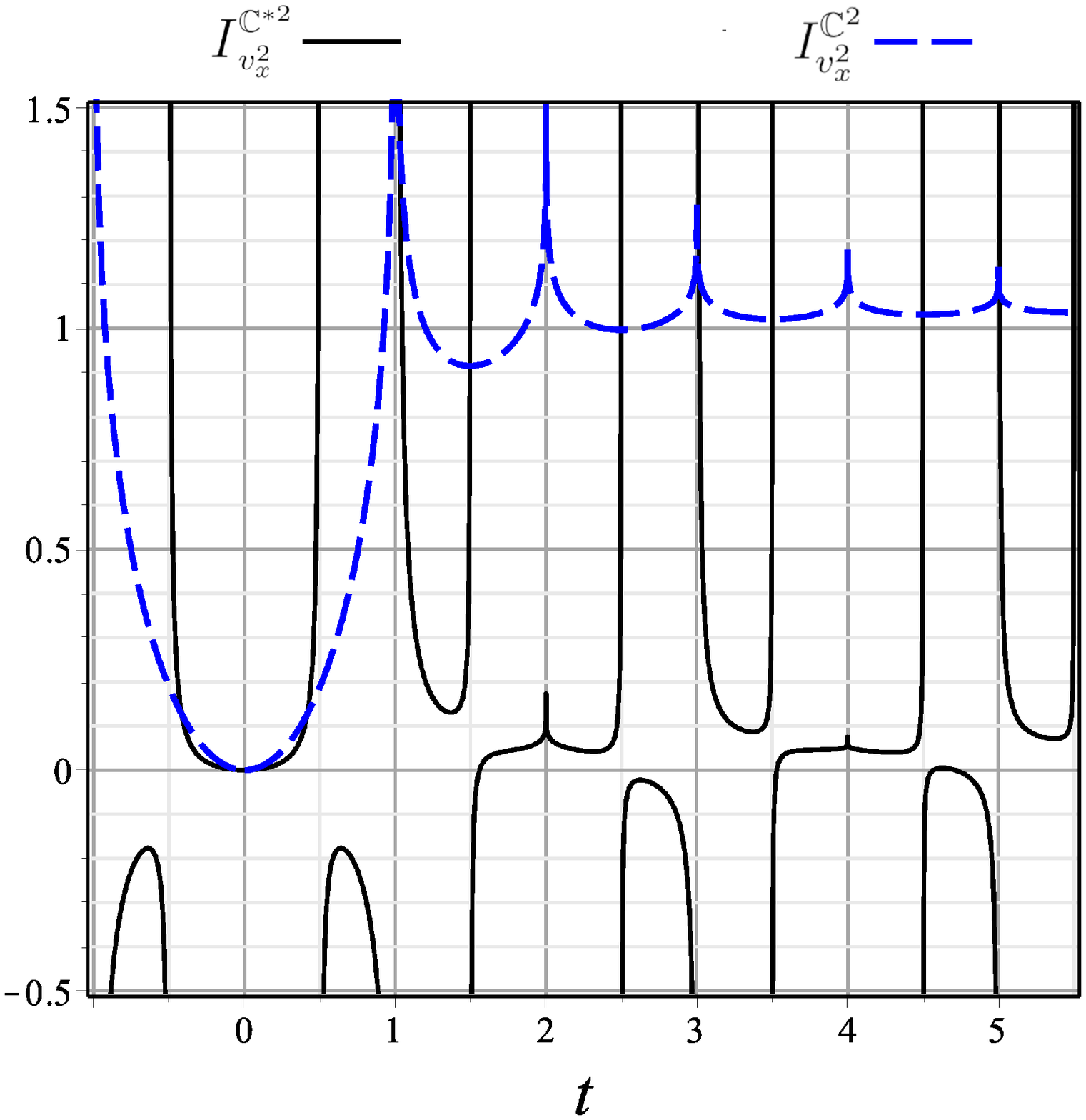}}    
    &    \resizebox{7cm}{!}{\includegraphics{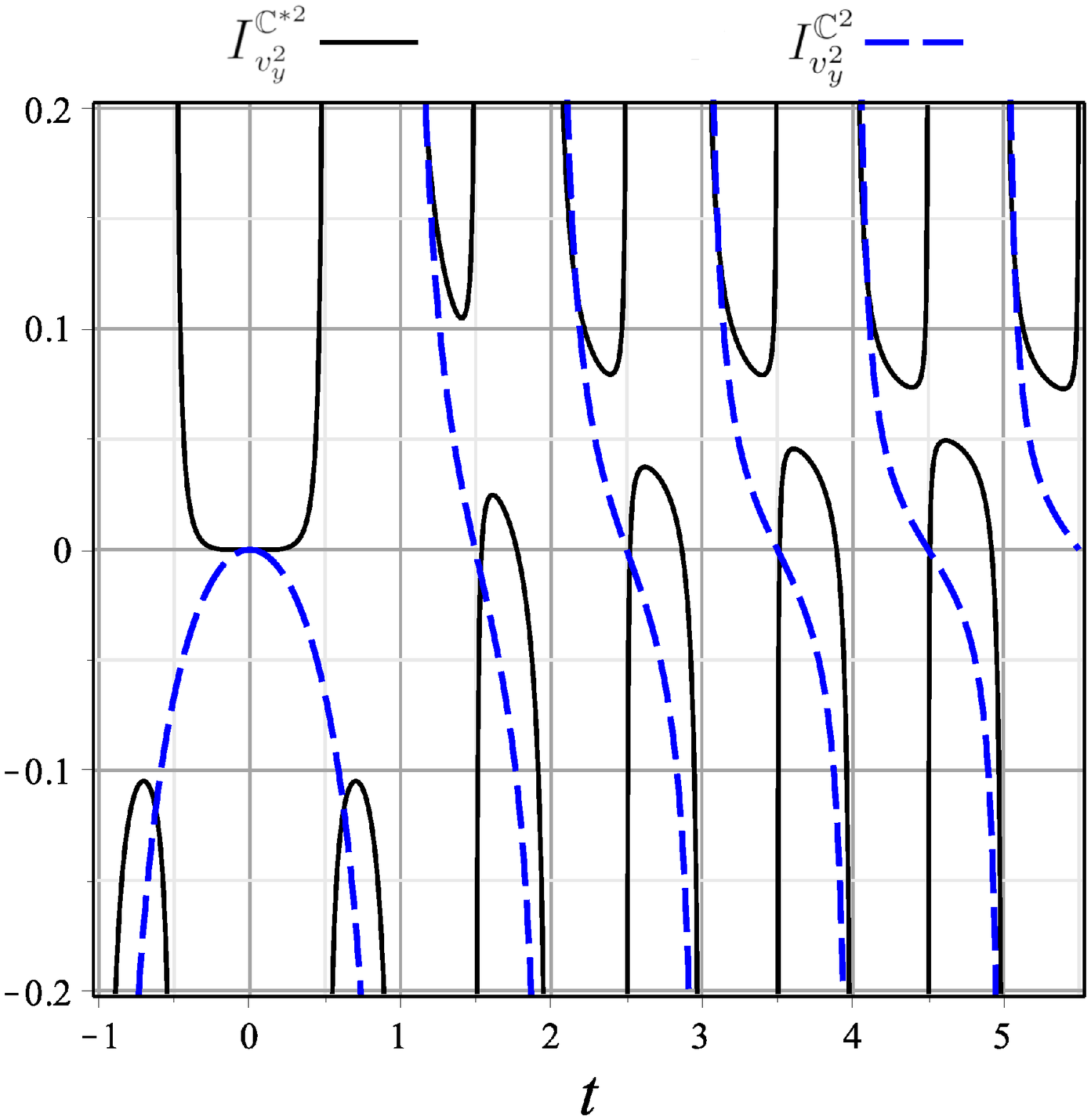}}  \\
       (a) & (b)
  \end{tabular}
    \end{center}
    \caption{Comparison of two components of the indicator for the twisted cylinder
    $\mathbb{C}^{*2} \times \mathbb{E}^2$ with those for its time orientable counterpart
    $\mathbb{C}^2 \times \mathbb{E}^2$. In both plots $t_0=0$. Panel (a) shows the
    $x$-component of the indicator for $\mathbb{C}^{*2} \times \mathbb{E}^2$  given
    by \eqref{indicator-i-TNS} plotted together with the $x$-component of the indicator
    for $\mathbb{C}^2 \times \mathbb{E}^2$  given by \eqref{IE16y}. Panel (b) shows the
    same comparison for the $y$-component of the indicator. In these plots, $q,\,m$ and $a$
    are all set to unity for qualitative reasons only. The solid curves for
    $\mathbb{C}^{*2} \times \mathbb{E}^2$ present an inversion pattern, roughly of the form $\cup$
followed by $\cap$, which is absent from the dashed curves for the time-orientable
$\mathbb{C}^2 \times \mathbb{E}^2$.
\label{Fig1}}
\end{figure*}

\begin{figure*}[htpb]
 \begin{center}
\begin{tabular}{c c}
\resizebox{7cm}{!}{\includegraphics{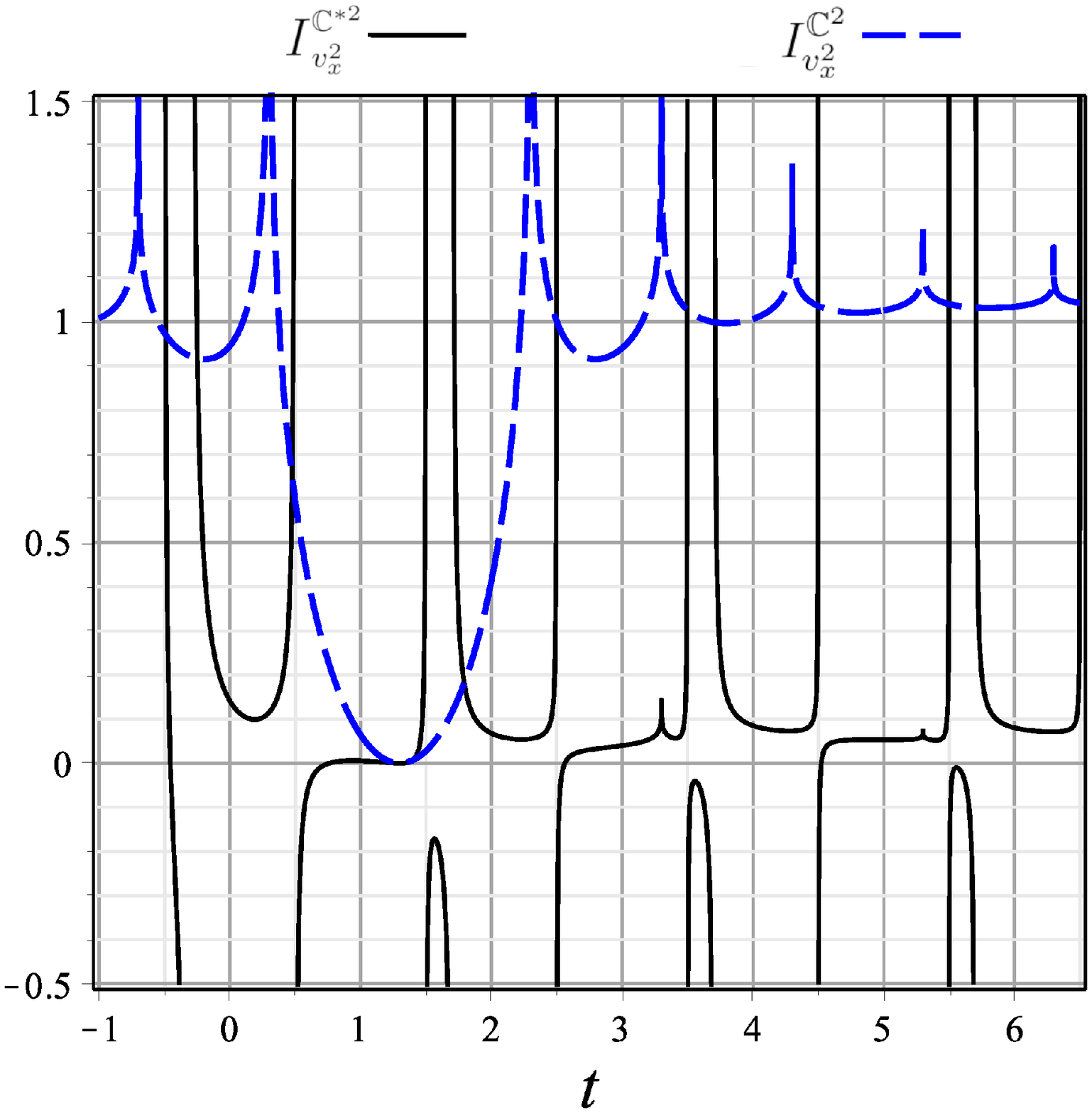}}    
    &    \resizebox{7cm}{!}{\includegraphics{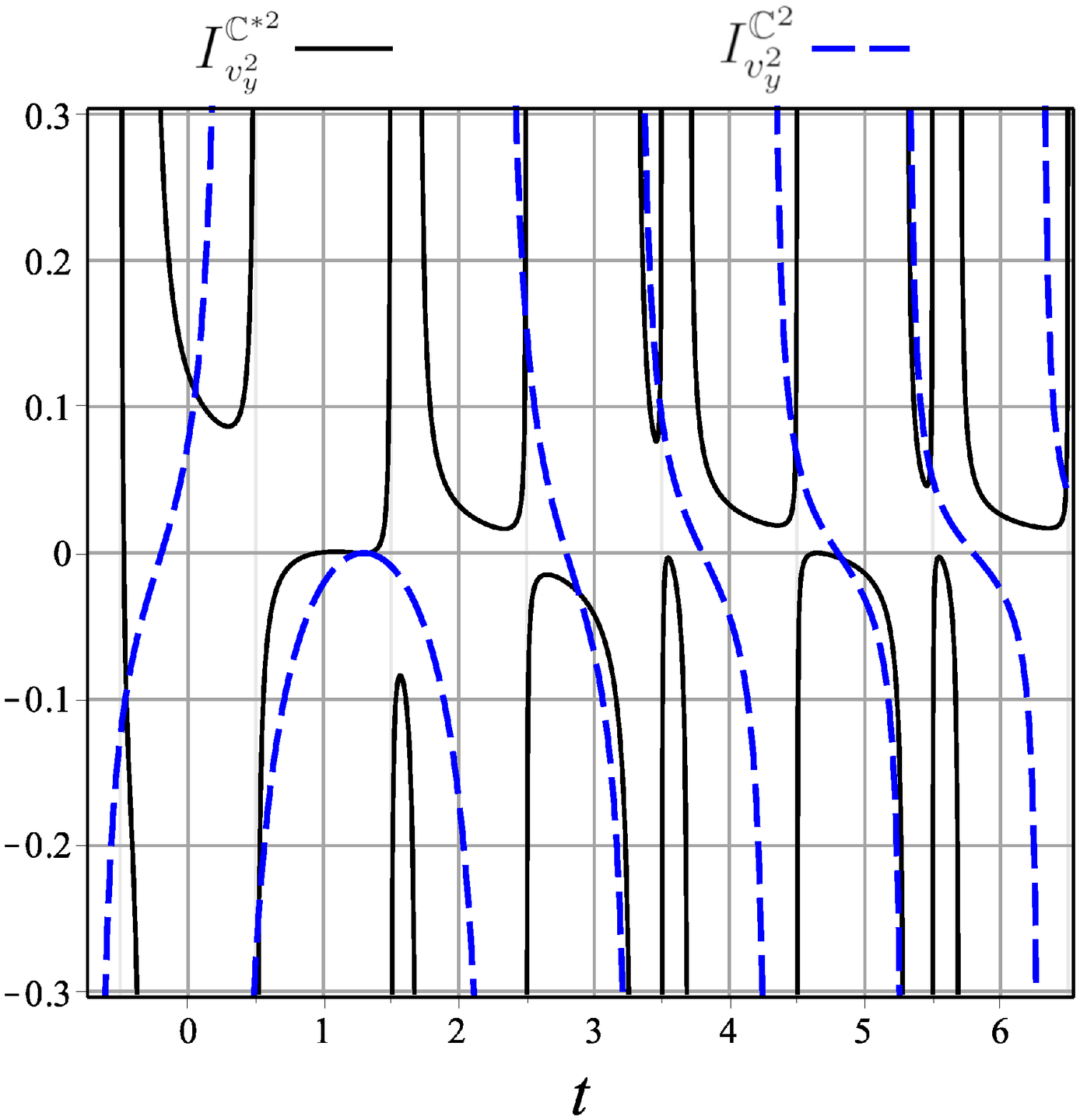}}  \\
       (a) & (b)
  \end{tabular}
    \end{center}
    \caption{The same as in Fig. \ref{Fig1} for $t_0=1.3$. Again, the solid curves
    for  $\mathbb{C}^{*2} \times \mathbb{E}^2$ exhibit an inversion pattern, roughly
    of the form $\cup$ followed by $\cap$, which is absent from dashed the curves
    for its time-orientable counterpart $\mathbb{C}^2 \times \mathbb{E}^2$.
\label{Fig2}}
\end{figure*}

Figure~\ref{Fig3}  illustrates quite vividly how the global inhomogeneity of
the twisted cylinder manifests itself through the statistical indicator~\eqref{indicator-i-TNS}.
Next we study how the indicator \eqref{indicator-i-TNS} for the twisted cylinder behaves
as a function of $t$ for fixed $t_0$.
Figures~\ref{Fig1} and~\ref{Fig2} compare components of the indicator for
$\mathbb{C}^{*2} \times \mathbb{E}^2$ with those for its time orientable counterpart
$\mathbb{C}^2 \times \mathbb{E}^2$ for two values of $t_0$. Both for $t_0=0$
(Fig.~\ref{Fig1}) and $t_0=1.3$ (Fig. \ref{Fig2}) there appears an inversion pattern
in the case of  $\mathbb{C}^{*2} \times \mathbb{E}^2$, roughly of the form $\cup$
followed by $\cap$. The absence of any inversion pattern in the case of  the
time-orientable $\mathbb{C}^2 \times \mathbb{E}^2$ allows one to pinpoint
the temporally non-orientable case.

In order to demonstrate our main result, namely the possible local detection of
time non-orientability, we have chosen two spacetime manifolds of the form
$\mathcal{M}_4=\mathcal{M}^2\times\mathbb{E}^2$ such that the first factors
are topologically similar in that  each is compact in just one direction and
has only one discrete  isometry generator $\gamma_i$ (see Section II).
Thus, the only difference in their construction is that for the cylinder the generator
$\gamma_i$ is a translation whereas for the twisted cylinder
it is a glide reflection (translation followed by an inversion or flip),
which suffices to make the quotient manifold non-orientable.
The repeated inversion pattern in the curves for indicator
$\mbox{\large $I$}_{v^2_i}^{\mathbb{C}^{*2}}$, Eq.~\eqref{indicator-TNS},
for the spacetime manifold $\mathcal{M}_4$  with the time non-orientable
twisted cylinder factor $\mathcal{M}^2 = \mathbb{C^*}^2$, constitutes an
exclusive signature of time-non-orientability.
As for the spacetime manifold $\mathcal{M}_4$ with the cylinder factor
$\mathcal{M}^2 = \mathbb{C}^2$, the periodic pattern of the curves is uniform
(shows no inversion) because $\gamma_i$ is a pure translation isometry.
The glide reflection is the only topological difference between the twisted cylinder
and the cylinder, therefore it is the cause of the different behaviors
of our orientability indicator in the two cases because the physical
system is the same.


\subsection{Possible physical realization}

A brief discussion of  the physical scale in which our  approach can be realized is fitting. From Eq. \eqref{indicator-i-TNS}, or more clearly from Eq. \eqref{IE16x}, it follows that the order of magnitude of the time-orientability indicators is
\begin{equation}
\label{Indicators-order-magnitude}
I \sim \frac{q^2 \Delta t}{4\pi m^2 a^3},
\end{equation}
where $a$ is the topological length scale and $\Delta t$ is the measurement duration.
The above equation holds in units such that $\hbar = c =1$. Let us consider the case in which the charged particle is an electron. Inserting the appropriate factors of $\hbar$ and $c$, and using $e^2/\hbar c = \alpha$ in CGS units, where $\alpha$ is the fine-structure constant, we  have
\begin{equation}
\label{Indicators-order-magnitude-h-c}
I \sim \frac{\alpha}{4\pi}\frac{\hbar^2 c\Delta t}{m^2 a^3},
\end{equation}
which has the correct dimension of velocity squared. Our analysis was performed for nonrelativistic charged particles. Accordingly, for the sake of estimation, let us take $I \sim (c/10^3)^2$. This gives
\begin{equation}
\label{Delta-t}
\Delta t \sim \frac{4\pi \times 137 m^2 c}{10^6 \hbar^2}\,a^3 \sim \bigg( 3.9 \times 10^{7} \frac{\mbox{s}}{\mbox{cm$^3$}}\bigg) a^3,
\end{equation}
where we have used  $\alpha \approx 1/137$, $c = 3 \times 10^{10}$ cm/s, $m= 9.1 \times 10^{-28}$ g and $\hbar = 1.05 \times 10^{-27}$ erg s. Consequently, if $a \lesssim 30 \,\mu\mbox{m} $ then
$\Delta t \lesssim 1$ s. This means that it would take no more that a second of observation to detect a possible time non-orientablity in mesoscopic or microscopic scale. The detection time increases proportionally to the cube of the topological length scale. Therefore, the present method would not be suitable for detection of a putative  time non-orientablity in the solar system  scale,  let alone in cosmological scale. Furthermore, for the early expanding  universe it is the  Friedmann-Lema\^{\i}tre-Robertson-Walker (FLRW)  geometry that must be considered,  rather than the Minkowski spacetime metric \cite{Lemos-Reboucas-2022}.

Our calculations imply that a time non-orientable nontrivial topology will give rise to a certain inversion pattern in the  time evolution curves of the statistical orientability indicator.  In the case of a time orientable nontrivial topology there will be no such inversion pattern. For  trivial topology (standard Minkowski spacetime) the indicator will vanish. In order to decide which of these alternatives is realized in nature, it is necessary to measure stochastic motions of charged particles induced by quantum electromagnetic  vacuum fluctuations.
Conceivably, experimental detection of such stochastic motions  might be feasible for confined electrons in a Paul trapp \cite{Matthiesen} or in a Penning trap \cite{Vogel}. It seems likely that  technical difficulties will have to be overcome  to tell apart the effects induced by quantum electromagnetic  vacuum fluctuations from those caused by the applied electromagnetic field that is responsible for  the confinement.

\section{Summary of results and final comments}  \label{Conclusions}

In relativistic cosmology and  quantum field theory,
 spacetime is assumed to be orientable as a topological manifold, and is additionally
endowed with a Lorentz metric, making it  separately time and space orientable.
The question as to whether the current laws of physics require that spacetime manifolds
adhere to these orientability assumptions are  among the unsettled  
issues in this framework.  
Although non-orientability is generally seen as an undesirable feature, 
non-orientable Lorentzian spacetime manifolds are concrete mathematical
possibilities  in physics at different scales. Therefore, the possibility that
spacetime is time non-orientable ought not to be jettisoned forthwith.
Strict orientability assumptions  could potentially
interfere with the understanding of fundamental aspects of the
interplay of physics and topology.
In fact, we do not know to what extent the topology of the underlying
spacetime manifold might encode, or whether it is necessary to capture or
express, basic features of the physical world in some scale. It is conceivable that
topological properties such as global homogeneity and orientability of
spacetime manifolds might be testable in both micro- and
macro-physical scales.
Previous speculations on time orientability violation have revolved about gedanken experiments
that would allegedly provide signatures of a putative time non-orientability of
spacetime~\cite{MarkHadley-2002,MarkHadley-2018}. These ideas and  other approaches
to testing time orientability have been recently subjected to a searching
critique~\cite{Bielinska}. Here, however, we have settled  on  probing the orientability
of spacetime by means of measurable local physical effects.

Inasmuch as the role played by time orientability  is more clearly accessed
in static flat spacetime, whose  dynamical degrees of freedom are frozen,
in the present paper, instead of the expanding
FLRW spacetime,
we have focused on the time orientability of Minkowski spacetime, leaving
for a forthcoming article~\cite{Lemos-Reboucas-2022} some important related
questions regarding the so-called arrow of time --- observed time asymmetry
in macrophysics and in the evolution of the universe, despite the
time reversal invariance of the fundamental laws of physics.
This includes, for example, whether temporal orientability of the FLRW universe
can be probed and whether one can understand in a far-reaching context
the several existing arrows of time~\cite{Ellis-Drosel-2020,Barbour-2000,Rovelli-2018}.

In this paper we have argued that a presumed time non-orientability of Minkowski
empty spacetime can be locally probed through  physical effects associated with
quantum vacuum electromagnetic fluctuations.
To this end, we have studied the stochastic motions of a charged particle
under these fluctuations in Minkowski spacetime manifold, $\mathcal{M}_4$,
endowed with a time non-orientable topology (the twisted cylinder
$\mathcal{M}_4 = \mathbb{C}^{*2}  \times \mathbb{E}^2$) and its time orientable
counterpart.
We have found that the  statistical topological indicator given by
Eq.~\eqref{indicator-TNS} is suitable to bring out the time non-orientability
in the twisted cylinder case. 
Accordingly, we have derived analytical expressions
for the statistical orientability indicator corresponding to
Minkowski spacetime manifold $\mathcal{M}_4$ equipped with time non-orienbtable
and its time orientable counterpart.

The chief conclusion of this work is reached through comparisons between
the stochastic motions of a charged test particle in Minkowski spacetime
endowed with each of the two spacetime topologies.
Since the expressions for the orientability indicators
 --- Eqs. \eqref{indicator-i-TNS}, \eqref{IE16x} and \eqref{IE16y} --- are
too involved for a direct comparison,
to demonstrate our main result, which is ultimately stated in
terms of patterns of curves for the orientability indicators,
we have performed numerical computations and  plotted figures for
the components of our statistical orientability indicator.

Figure~\ref{Fig3} aims to illustrate the  topological temporal
inhomogeneity for $\mathcal{M}_4$ corresponding to the twisted cylinder
$\mathbb{C}^{*2} \times \mathbb{E}^2$, a feature not shared by the cylinder
$\mathbb{C}^{2} \times \mathbb{E}^2$.
The patterns encountered are different for each fixed $t_0$, and the closed
contour curves are also very different from those for $\mathbb{C}^2 \times \mathbb{E}^2$,
which are straight $45$-degree lines. The topological singularity structure is also
markedly richer than the one for the time orientable case.

However, the answer to the central question of the paper, namely how to locally
probe the time orientability of Minkowski spacetime intrinsically, is accomplished
by comparing the time evolution of the  orientability indicators
$\mbox{\large $I$}_{v^2_i}^{\mathbb{C}^{*2}}$ and
$\mbox{\large $I$}_{v^2_i}^{\mathbb{C}^{2}}$.
Figs. \ref{Fig1} and \ref{Fig2} show
that it may be possible to locally  unveil time non-orientability through
the inversion pattern of curves of the time non-orientability indicator
for a charged particle under quantum vacuum electromagnetic
fluctuations. This inversion pattern is a signature of time
non-orientability and is absent in the time orientable case.

It should be stressed that the \textsl{time non-orientability} for the  twisted cylinder
is of topological origin (periodic inversion of time controlled by the topological length
$a$, the  fundamental domain size).
It is different from the continuous closed timelike curves  that come
about  in G\"odel~\cite{Godel} and other spacetime solutions of Einstein's equations
 \cite{CTLC-examples,CTLC-examples-2,CTLC-examples-3,CTLC-examples-4,CTLC-examples-5}.
Particularly, in the G\"odel model  the underlying manifold is topologically equivalent to
$\mathbb{R}^4$, there is no periodic time inversion of topological origin, but
gravity tilts the local light cones so as to allow continuous closed timelike curves  in the topologically trivial spacetime.

To summarize, the main result of our analysis is that it may be feasible to
look into a conceivable topological time non-orientability of  Minkowski empty spacetime by
measurable local physical effects associated with quantum vacuum
electromagnetic fluctuations.


\begin{acknowledgements}
M.J. Rebou\c{c}as acknowledges the support of FAPERJ under a CNE E-26/202.864/2017 grant,
and thanks CNPq for the grant under which this work was carried out.
We are grateful to A.F.F. Teixeira for a careful reading of the manuscript and
 discussions. 
\end{acknowledgements}


\end{document}